\documentclass[11pt]{article}

\usepackage{times,cite}

\newtheorem{Theorem}{Theorem}[section]
\newtheorem{Lemma}[Theorem]{Lemma}
\newtheorem{Definition}[Theorem]{Definition}
\newtheorem{Remark}[Theorem]{Remark}
\newtheorem{Corollary}[Theorem]{Corollary}

\usepackage[]{graphicx}

\usepackage{amsfonts}
\usepackage{url}
\usepackage[breaklinks,hidelinks]{hyperref}
\usepackage{bussproofs}

\newcommand{\text}[1]{\mbox{#1}}

\newcommand{\eqdef}{:\equiv}
\newcommand{\allst}{\forall^{\mathrm{st}}}
\newcommand{\exst}{\exists^{\mathrm{st}}}
\newcommand{\N}{\mathbb{N}}
\newcommand{\st}{\mathsf{st}}
\newcommand{\apl}[1]{[#1]}
\newcommand{\br}{\mathbf{r}}
\newcommand{\bt}{\mathbf{t}}
\newcommand{\bu}{\mathbf{u}}
\newcommand{\bv}{\mathbf{v}}
\newcommand{\el}{\varepsilon}
\newcommand{\bdf}{\mathbf{f}}
\newcommand{\bdg}{\mathbf{g}}

\def\Proof{\par \noindent{\bf Proof: }}
\def\Done{\hfill\rule{0.5em}{0.5em}}

\newcommand{\hfi}[3]{\{#1\}^{#2}_{#3}}
\newcommand{\afi}[3]{|#1|^{#2}_{#3}}

\newcommand{\pup}[2]{#1^{#2^+ \uparrow}}
\newcommand{\pdown}[2]{#1^{#2^+ \downarrow}}
\newcommand{\nup}[2]{#1^{#2^- \uparrow}}
\newcommand{\ndown}[2]{#1^{#2^- \downarrow}}

\newcommand{\IHarrow}{\stackrel{\small{\textup{IH}}}{\Rightarrow}}
\newcommand{\tin}{\,\mathord{\in}\,}
\newcommand{\ty}{\mathbf{\nu}}
\newcommand{\typ}{\mathbf{\delta}}

\newcommand{\HAomega}{\mathrm{E\textup{-}HA}^\omega}
\newcommand{\PAomega}{\mathrm{E\textup{-}PA}^\omega}
\newcommand{\stHAomega}{\textrm{E-HA}_{\st}^\omega}
\newcommand{\stPAomega}{\textrm{E-PA}_{\st}^\omega}
\newcommand{\NCR}[1]{\textrm{NCR}_{#1}}
\newcommand{\US}[1]{\textrm{US}_{#1}}
\newcommand{\TRall}[1]{\textrm{TR}^{\forall}_{#1}}
\newcommand{\TRex}[1]{\textrm{TR}^{\exists}_{#1}}

\newcommand{\hmr}[2]{#2 \; \textrm{hr} \; #1}

\begin{document}

\title{On the Herbrand Functional Interpretation}

\author{Paulo Oliva and Chuangjie Xu}

\maketitle

\begin{abstract} We show that the types of the witnesses in the Herbrand functional interpretation can be simplified, avoiding the use of ``sets of functionals" in the interpretation of implication and universal quantification. This is done by presenting an alternative formulation of the Herbrand functional interpretation, which we show to be equivalent to the original presentation. As a result of this investigation we also strengthen the monotonicity property of the original presentation, and prove a monotonicity property for our alternative definition.
\end{abstract}

\section{Introduction}

Nonstandard Analysis, pioneered by Abraham Robinson \cite{Robinson(66)}, gives a formal treatment of the intuitive ``calculus with infinitesimals'' common in physics and mathematics until Weierstrass' epsilon-delta framework became widespread. Nelson's internal set theory \cite{Nelson(77)} is an axiomatic approach to Nonstandard Analysis based on ZFC enriched with a predicate ``$x$ is standard". Nelson's framework has been formulated for higher-order Peano and Heyting arithmetic, and given a computational interpretation via Herbrand realizability and the associated nonstandard Dialectica interpretation \cite{Berg(2012)} -- we refer to the later here as the Herbrand functional interpretation. More recently, this Herbrand functional interpretation has been successfully applied in the characterisation of the computational content of nonstandard analysis \cite{Sanders(16),Sanders(17),Sanders(18)}. 

Basic nonstandard Peano and Heyting arithmetic in all finite types, denoted $\stHAomega$ and $\stPAomega$, are obtained by extending the language of $\HAomega$ and $\PAomega$ with a new type construct $\sigma^*$, for finite non-empty subsets of objects of type $\sigma$, and a new family of predicate symbols $\st^\sigma(x)$ -- denoting that $x$ is a \emph{standard} element of type $\sigma$. These basic systems are then enriched with higher-type generalisations of principles from nonstandard analysis \cite{Nelson(77),Robinson(66)} such as the \emph{realization principle}
\[ 
\NCR{\tau, \sigma} \quad \colon \quad 
	\forall y^\tau \exst x^\sigma A(x, y) \to \exst x^{\sigma^*} \forall y^\tau \exists x' \tin x A(x', y),
\]
\emph{underspill}
\[ 
\US{\sigma} \quad \colon \quad 
	\forall x^\sigma (\neg \st^\sigma(x) \to A(x)) \to \exst x^\sigma A(x),
\]
and the \emph{transfer rules}
\[
    \AxiomC{$\allst x^\tau A(x)$}
    \RightLabel{$(\TRall{\tau})$}
    \UnaryInfC{$\forall x^\tau A(x)$}
    \DisplayProof
    \quad \quad
    \AxiomC{$\exists x^\tau A(x)$}
    \RightLabel{$(\TRex{\tau})$}
    \UnaryInfC{$\exst x^\tau A(x)$}
    \DisplayProof.
\]
The key novelty of these interpretations is the realization of the new predicate symbol $\st^\sigma(x)$ via finite subsets of objects of type $\sigma$, i.e.\ one says that $s^{\sigma^*}$ realises $\st^\sigma(x)$ whenever $x \in s$. As a consequence of this the new standardness predicate is eliminated, but an existential quantification over standard elements is realised by a set of potential witnesses, very much like in a Herbrand disjunction, i.e.
\[ \hmr{\exst x^\sigma A}{s, \br} \quad \eqdef \quad \exists x \tin s (\hmr{A}{\br}) \]
where we use bold face letters such as $\br$ to denote tuples of variables or terms.

In fact, in the \emph{nonstandard} interpretations (both Herbrand realizability and functional interpretation), the realisers of all formulas are of type $\sigma^*$, for some type $\sigma$, i.e.\ the realisers are always finite sets. This was done so that one can demonstrate the \emph{monotonicity property}
\[ \text{if } \hmr{A}{\br} \text{ and } \br \subseteq \br' \text{ then } \hmr{A}{\br'} \]
so if a sequence of sets $\br$ realises a formula $A$, then any sequence of larger sets will also realise $A$. But that introduces a so-called challenge in the interpretations of $A \to B$ and $\allst x^\sigma A$, which are normally interpreted via \emph{functionals}, not \emph{sets}!

The ingenuous solution in \cite{Berg(2012)} was to consider a new form of application operation on sets of functions. Given a set of functionals $\bdf \colon (\mathbf{\sigma}^* \to \mathbf{\tau}^*)^*$ and a set of arguments $\br \colon \mathbf{\sigma}^*$, one can ``apply'' $\bdf$ to $\br$ as
\[ \bdf \apl{\br} \quad \eqdef \quad \bigcup_{\bdf' \in \bdf} \bdf'(\br) \]
so that the interpretations of implication $A \to B$ and universal quantification $\allst x^\sigma A$ are defined using this novel form of \emph{set application}, e.g.
\[ \hmr{A \to B}{\bdf} \quad \equiv \quad \forall \br ( \hmr{A}{\br} \to \hmr{B}{\bdf \apl{\br} }). \]

It has been recently shown (cf. \cite{Oliva(2019)}), however, that an alternative definition of the Herbrand realizability is possible, where this new \emph{set application} is avoided, and the usual functional application is used instead. The main purpose of the present note is to show that the same holds for the Herbrand functional interpretation: the use of the \emph{set application} is not essential. We present an alternative formulation of the Herbrand functional interpretation (Definitions \ref{def-down-types} and \ref{def-afi}) which uses the standard functional application. The two formulations are shown to be equivalent (Theorem \ref{main-thm}) via constructions (Definition \ref{up-down-constructions}) that transform realisers for one presentation into another. As a byproduct of our investigation on the monotonicity of the new presentation (Lemma \ref{mon-strong}), we have also discovered that the original Herbrand functional interpretation satisfies a stronger monotonicity property (Lemma \ref{monotonicity2}) than originally thought \cite[Lemma~5.4]{Berg(2012)}, since we can prove monotonicity with respect to a relation which is weaker than subset inclusion. Throughout the paper we will use $A \equiv \allst n^\N \exst m^\N ( n = m )$ as our running example to illustrate the definitions and constructions.


Since the focus of this paper is more on the structure of the Herbrand functional interpretation itself, rather than the fact that it interprets $\stHAomega$, we will refer the readers to \cite{Berg(2012)} for the definitions of the systems $\stHAomega$ and $\stPAomega$. We just remark that, as in \cite{Berg(2012)}, we will be taking $\allst x^\sigma A$ and $\exst x^\sigma A$ to be primitive notions axiomatised as
\[ 
\allst x^\sigma A \Leftrightarrow \forall x^\sigma (\st^\sigma(x) \to A)
\qquad \qquad
\exst x^\sigma A \Leftrightarrow \exists x^\sigma (\st^\sigma(x) \wedge A).
\]
This is done in order to take advantage of the fact that the interpretation of $$\forall x^\sigma (\st^\sigma(x) \to A)$$ can be simplified once one observes the monotonicity property mentioned above (cf. Remark \ref{remark-allst}). For simplicity, we will also restrict ourselves to the fragment without disjunction $A \vee B$, since in the theories above disjunction can be defined from the other connectives and quantifiers as
\[ A \vee B \equiv \exists n^\N ((n = 0 \to A) \wedge (n \neq 0 \to B)). \]
{\bf Acknowledgements}. We would like to thank the anonymous referee for several helpful suggestions.

\section{The Herbrand Functional Interpretation}

Similarly to G\"odel's original functional (Dialectica) interpretation \cite{Goedel(58)}, the Herbrand functional interpretation maps each formula $A$ to a new formula $A_{D_{\st}}(\br, \bu)$ with two extra sets of free-variables $\br$ and $\bu$. In the Herbrand functional interpretation $A$ will be in the language of $\stHAomega$ whereas $A_{D_{\st}}(\br, \bu)$ will be a formula of $\HAomega$. We think of $\br$ as the \emph{positive} information in $A$, and of $\bu$ as the \emph{negative} information in $A$, since $A$ can be shown, over an appropriate system, to be equivalent to $\exists \br \forall \bu A_{D_{\st}}(\br, \bu)$. In order to better distinguish between the positive and negative information of $A$, we will from now on write $A_{D_{\st}}(\br, \bu)$ as $\hfi{A}{\br}{\bu}$. Let us start by determining the types of these tuples $\bu$ and $\br$:

\begin{Definition}[$\ty$-types] \label{up types} To each formula $A$ of $\stHAomega$ let us associate two tuples $\ty^+_A$ and $\ty^-_A$ of types inductively in the structure of $A$ as follows
\[
{\arraycolsep=2pt
\def\arraystretch{1.3}
\begin{array}{rclcrcl}
\ty^+_{\st^\sigma(t)} & \eqdef & \sigma^* & \hspace{10mm} &
\ty^-_{\st^\sigma(t)} & \eqdef & \varepsilon \\
\ty^+_{A \wedge B} & \eqdef & \ty^+_A, \ty^+_B & &
\ty^-_{A \wedge B} & \eqdef & \ty^-_A, \ty^-_B \\
\ty^+_{A \to B} & \eqdef & (\ty^+_A \to \ty^+_B)^* , (\ty^+_A \to \ty^-_B \to (\ty^-_A)^*)^* & &
\ty^-_{A \to B} & \eqdef & \ty^+_A, \ty^-_B \\
\ty^+_{\forall x^\sigma A} & \eqdef & \ty^+_A & &
\ty^-_{\forall x^\sigma A} & \eqdef & \ty^-_A \\
\ty^+_{\exists x^\sigma A} & \eqdef & \ty^+_A & &
\ty^-_{\exists x^\sigma A} & \eqdef & (\ty^-_A)^* \\
\ty^+_{\allst x^\sigma A} & \eqdef & (\sigma \to \ty^+_A)^* & &
\ty^-_{\allst x^\sigma A} & \eqdef & \sigma, \ty^-_A \\
\ty^+_{\exst x^\sigma A} & \eqdef & \sigma^*, \ty^+_A & &
\ty^-_{\exst x^\sigma A} & \eqdef & (\ty^-_A)^* \\[2pt]
\end{array}}
\]
\end{Definition}

While the atomic formulas $\st^\sigma(t)$ will have computational content, the atomic formulas $s =_{\N} t$ are ``computationally empty". Therefore, for the rest of the paper we will only consider the atomic formulas $\st^\sigma(t)$ (and omit the atomic formulas $s = t$) when defining constructions or proving results by induction on the logical structure of a formula. In each of these, the base case when $A$ is $s = t$ will be trivial.

We will refer to the $\ty$-types as the \emph{up types} -- in contrast to the types of Definition \ref{def-down-types} which we will call the \emph{down types}. Note that the positive up types $\ty^+_{A}$ are all of the form $\sigma_1^*, \ldots, \sigma_n^*$ for some $\sigma_1, \ldots, \sigma_n$. For instance, considering our running example $A \equiv \allst n^\N \exst m^\N ( n = m )$, according to Definition \ref{up types}, we have that $\ty^+_A \equiv (\N \to \N^*)^*$ and $\ty^-_A \equiv \N$.

\begin{Definition}[Herbrand functional interpretation, {\cite[Definition~5.1]{Berg(2012)}}] \label{def-hfi} For each formula $A$ in the language of $\stHAomega$ associate a formula $\hfi{A}{\br}{\bu}$, in the language of $\HAomega$, by induction on $A$, where $\br : \ty^+_A$ and $\bu : \ty^-_A$, as follows:
\[
{\arraycolsep=2pt
\def\arraystretch{1.5}
\begin{array}{rcl}
    \hfi{\st^\sigma(z)}{s}{\el} & \eqdef & z \in s \\
    \hfi{A \wedge B}{\br,\bt}{\bu,\bv} 
    	& \eqdef & \hfi{A}{\br}{\bu} \wedge \hfi{B}{\bt}{\bv} \\
    \hfi{A \to B}{\bdf,\bdg}{\br,\bu} 
    	& \eqdef & \forall \bv \tin \bdg\apl{\br,\bu} \hfi{A}{\br}{\bv} \to \hfi{B}{\bdf\apl{\br}}{\bu} \\
    \hfi{\forall x^\sigma A(x)}{\br}{\bu} & \eqdef 
    	& \forall x^\sigma \hfi{A(x)}{\br}{\bu} \\
    \hfi{\exists x^\sigma A(x)}{\br}{\bu} & \eqdef 
    	& \exists x^\sigma \, \forall \bv \tin \bu \hfi{A(x)}{\br}{\bv} \\
    \hfi{\allst x^\sigma A(x)}{\bdf}{c,\bu} 
    	& \eqdef & \hfi{A(c)}{\bdf\apl{c}}{\bu} \\
    \hfi{\exst x^\sigma A(x)}{s,\br}{\bu} 
    	& \eqdef & \exists x \tin s \, \forall \bv \tin \bu \hfi{A(x)}{\br}{\bv} 
\end{array}}
\]
\end{Definition}

Making use of our running example $A \equiv \allst n^\N \exst m^\N ( n = m )$ we see that $\hfi{A}{f}{n} \equiv \exists m \tin f \apl{n} (n = m)$, where $f \colon (\N \to \N^*)^*$ and $n \colon \N$.

If $\mathbf{\sigma} = \sigma_1^*, \ldots, \sigma_n^*$ and $\br,\br' : \mathbf{\sigma}$, then we write $\br \subseteq_{\mathbf{\sigma}} \br'$ as an abbreviation for $r_1 \subseteq_{\sigma_1^*} r'_1 \wedge \ldots \wedge r_n \subseteq_{\sigma_n^*} r'_n$.

\begin{Lemma}[Monotonicity, {\cite[Lemma~5.4]{Berg(2012)}}] \label{monotonicity} $\hfi{A}{\br}{\bu} \wedge \br \subseteq_{\ty^+_A} \br' \vdash_{\HAomega} \hfi{A}{\br'}{\bu}$.
\end{Lemma}

\begin{Remark} \label{remark-allst} Given that $A$ has interpretation $\hfi{A}{\br}{\bu}$, we note that the interpretation of $\forall x^\sigma (\st^\sigma(x) \to A)$ is
\[
\hfi{\forall x^\sigma (\st^\sigma(x) \to A)}{\bdf}{s,\bu} \equiv \forall c \tin s \, \hfi{A(c)}{\bdf\apl{s}}{\bu}.
\]
The fact that this interpretation can be simplified to
\[
\hfi{\allst x^\sigma A(x)}{\bdf}{c,\bu} \equiv \hfi{A(c)}{\bdf\apl{c}}{\bu}
\]
relies on the monotonicity property above. More specifically, monotonicity is used to prove the soundness of 
\[ \allst x^\sigma A \vdash \forall x^\sigma (\st^\sigma(x) \to A). \]
\end{Remark}

Let us observe that a strengthening of Lemma \ref{monotonicity} is possible. We can motivate this strengthening using our running example $A \equiv \allst n^\N \exst m^\N ( n = m )$. Recall that $\ty^+_A \equiv (\N \to \N^*)^*$ and $\hfi{A}{f}{n} \equiv \exists m \tin f \apl{n} (n = m)$. It is clear that $t^{\ty^+_A} \equiv \{ \lambda n.\{ n \} \}$ witnesses $A$, since
\[ 
\hfi{A}{\{ \lambda n'.\{ n' \} \}}{n} 
\quad \equiv \quad  
\exists m \tin \{ \lambda n'.\{ n' \} \} \apl{n} (n = m) 
\quad \Leftrightarrow \quad 
\exists m \tin \{n\} (n = m).
\]
But so do $t_1^{\ty^+_A} \equiv \{ \lambda n.\{ n \}, \lambda n.\{ n + 1 \} \}$ and $t_2^{\ty^+_A} \equiv \{ \lambda n.\{ n, n + 1 \} \}$. The fact that $t_1$ is also a witnesses follows directly from Lemma \ref{monotonicity}, since $t \subseteq_{\ty^+_A} t_1$. But we do not have $t \subseteq_{\ty^+_A} t_2$, although it is clear that $t_2$ is a ``weakening" of $t$, in the sense that the function $\lambda n . \{ n \}$ is replaced by the weaker function $\lambda n . \{ n , n + 1 \}$. The following definition introduces a partial order on the positive up types $\ty^+_A$ which is weaker than the subset relation:

\begin{Definition} Define the following partial order $\br \sqsubseteq_A \br'$ between $\br, \br' \colon \ty^+_A$ by induction on $A$ as follows:
\[
{\arraycolsep=2pt
\def\arraystretch{1.5}
\begin{array}{rcl}
    s \sqsubseteq_{\st^\sigma(z)} s' 
    	& \eqdef & s \subseteq_{\sigma^*} s' \\
    \br, \bt \sqsubseteq_{A \wedge B} \br', \bt' 
    	& \eqdef & (\br \sqsubseteq_A \br') \wedge (\bt \sqsubseteq_B \bt') \\
    \bdf, \bdg \sqsubseteq_{A \to B} \bdf', \bdg' 
    	& \eqdef & \forall \br (\bdf \apl{\br} \sqsubseteq_B \bdf' \apl{\br} ) \wedge
			\forall \br, \bu (\bdg \apl{\br, \bu} \subseteq_{(\ty^-_A)^*} \bdg' \apl{\br, \bu})  \\
    \br \sqsubseteq_{\forall x^\sigma A} \br' 
    	& \eqdef & \br \sqsubseteq_{A} \br' \\
    \br \sqsubseteq_{\exists x^\sigma A} \br' 
    	& \eqdef & \br \sqsubseteq_{A} \br' \\
    \bdf \sqsubseteq_{\allst x^\sigma A} \bdf' 
    	& \eqdef & \forall c^\sigma (\bdf \apl{c} \sqsubseteq_{A} \bdf' \apl{c}) \\
    s, \br \sqsubseteq_{\exst x^\sigma A} s', \br' 
    	& \eqdef & (s \subseteq_{\sigma^*} s') \wedge (\br \sqsubseteq_{A} \br')
\end{array}}
\]
\end{Definition}

Note that in the definitions for $\bdf, \bdg \sqsubseteq_{A \to B} \bdf', \bdg'$ and $\bdf \sqsubseteq_{\allst x^\sigma A} \bdf'$ we are not comparing the sets of functionals themselves, but rather comparing the sets of values these functionals produce.

\begin{Lemma} \label{lemma-order} Let $\br, \br' \colon \ty^+_A$.
\begin{enumerate}
	\item If $\br \subseteq_{\ty^+_A} \br'$ then $\br \sqsubseteq_A \br'$.
	\item There are $\br$ and $\br'$ such that  $\br \sqsubseteq_A \br'$ but not $\br \subseteq_{\ty^+_A} \br'$.
\end{enumerate}
\end{Lemma}
\Proof
1. By induction on $A$. We show here only the case of implication. Assuming $\bdf,\bdg \subseteq_{\ty^+_{A \to B}} \bdf',\bdg'$ we have $\bdf\apl{\br} \,\subseteq_{\ty^+_B}\, \bdf'\apl{\br}$ and $\bdg\apl{\br,\bu} \,\subseteq_{(\ty^-_A)^*}\, \bdg'\apl{\br,\bu}$, for any $\br^{\ty^+_A}$ and $\bu^{\ty^-_B}$. By the induction hypothesis we get $\bdf\apl{\br} \,\sqsubseteq_{\ty^+_B}\, \bdf'\apl{\br}$ and hence $\bdf,\bdg \sqsubseteq_{A \to B} \bdf',\bdg'$. \\
2. We can use here our running example $A \equiv \allst n^\N \exst m^\N (n=m)$ with $\ty^+_A \equiv (\N \to \N^*)^*$. Let $t \eqdef \{ \lambda n.\{n\}, \lambda n.\{n+1\}\}$ and $t' \eqdef \{ \lambda n. \{ n, n+1 \} \}$. We have neither $t \subseteq_{\ty^+_A} t'$ nor $t' \subseteq_{\ty^+_A} t$, but we have that $t \sqsubseteq_{A} t'$ and $t' \sqsubseteq_{A} t$.
\Done

\begin{Lemma}[Strengthening of Lemma \ref{monotonicity}] \label{monotonicity2} $\hfi{A}{\br}{\bu} \wedge \br \sqsubseteq_A \br' \vdash_{\HAomega} \hfi{A}{\br'}{\bu}$.
\end{Lemma}

\Proof By induction on $A$. We show the case of implication. Assume $\hfi{A \to B}{\bdf,\bdg}{\br,\bu}$ and $\bdf,\bdg \sqsubseteq_{A \to B} \bdf',\bdg'$. We have $\bdf \apl{\br} \sqsubseteq_B \bdf' \apl{\br}$ and $\bdg\apl{\br,\bu} \subseteq_{(\ty^-_A)^*} \bdg'\apl{\br,\bu}$, and hence
\[
\begin{array}{lcl}
\hfi{A \to B}{\bdf,\bdg}{\br,\bu}
& \equiv &
\forall \bv \tin \bdg\apl{\br,\bu} \hfi{A}{\br}{\bv} \to \hfi{B}{\bdf\apl{\br}}{\bu} \\[2mm]
& \IHarrow &
\forall \bv \tin \bdg'\apl{\br,\bu} \hfi{A}{\br}{\bv} \to \hfi{B}{\bdf'\apl{\br}}{\bu} \\[2mm]
& \equiv &
\hfi{A \to B}{\bdf',\bdg'}{\br,\bu}. 
\end{array}
\]
The fact that the above lemma is a strengthening of Lemma \ref{monotonicity} follows from Lemma~\ref{lemma-order}.
\Done

\section{Alternative Presentation of the Herbrand Functional Interpretation}

Our main goal is to show that the set application used in the treatment of $A \to B$ and $\allst x^\sigma A$ can be replaced with the usual functional application. This will mean that the positive types are no longer of the form $\sigma_1^*, \ldots, \sigma_n^*$, but might contains functionals $\sigma \to \tau$. For that reason we define the \emph{down types}:

\begin{Definition}[$\typ$-types] \label{def-down-types} To each formula $A$ of $\stHAomega$ let us associate two tuples $\typ^+_A$ and $\typ^-_A$ of types inductively in the structure of $A$ as follows:
\[
{\arraycolsep=2pt
\def\arraystretch{1.3}
\begin{array}{rclcrcl}
\typ^+_{\st^\sigma(t)} & \eqdef & \sigma^* & \qquad &
\typ^-_{\st^\sigma(t)} & \eqdef & \varepsilon \\
\typ^+_{A \wedge B} & \eqdef & \typ^+_A, \typ^+_B & &
\typ^-_{A \wedge B} & \eqdef & \typ^-_A, \typ^-_B \\
\typ^+_{A \to B} & \eqdef & (\typ^+_A \to \typ^+_B) , (\typ^+_A \to \typ^-_B \to (\typ^-_A)^*) & &
\typ^-_{A \to B} & \eqdef & \typ^+_A, \typ^-_B \\
\typ^+_{\forall x^\sigma A} & \eqdef & \typ^+_A & &
\typ^-_{\forall x^\sigma A} & \eqdef & \typ^-_A \\
\typ^+_{\exists x^\sigma A} & \eqdef & \typ^+_A & &
\typ^-_{\exists x^\sigma A} & \eqdef & (\typ^-_A)^* \\
\typ^+_{\allst x^\sigma A} & \eqdef & \sigma \to \typ^+_A & &
\typ^-_{\allst x^\sigma A} & \eqdef & \sigma, \typ^-_A \\
\typ^+_{\exst x^\sigma A} & \eqdef & \sigma^*, \typ^+_A & &
\typ^-_{\exst x^\sigma A} & \eqdef & (\typ^-_A)^* \\[2pt]
\end{array}}
\]
We will refer to the $\typ$-types as \emph{down types}.
\end{Definition}

We are now ready to present our alternative definition of the Herbrand functional interpretation:

\begin{Definition}[Alternative presentation of the Herbrand functional interpretation] \label{def-afi} For each formula $A$ in the language of $\stHAomega$ associate a formula $\afi{A}{\br}{\bu}$, in the language of $\HAomega$, by induction on $A$, where $\br : \typ^+_A$ and $\bu : \typ^-_A$, as follows:
\[
{\arraycolsep=2pt
\def\arraystretch{1.5}
\begin{array}{rcl}
    \afi{\st^\sigma(z)}{s}{\el} & \eqdef & z \in s \\
    \afi{A \wedge B}{\br,\bt}{\bu,\bv} 
    	& \eqdef & \afi{A}{\br}{\bu} \wedge \afi{B}{\bt}{\bv} \\
    \afi{A \to B}{\bdf,\bdg}{\br,\bu} 
    	& \eqdef & \forall \bv \tin \bdg \br \bu \afi{A}{\br}{\bv} \to \afi{B}{\bdf \br}{\bu} \\
    \afi{\forall x^\sigma A(x)}{\br}{\bu} & \eqdef 
    	& \forall x^\sigma \afi{A(x)}{\br}{\bu} \\
    \afi{\exists x^\sigma A(x)}{\br}{\bu} & \eqdef 
    	& \exists x^\sigma \, \forall \bv \tin \bu \afi{A(x)}{\br}{\bv} \\
    \afi{\allst x^\sigma A(x)}{\bdf}{c,\bu} 
    	& \eqdef & \afi{A(c)}{\bdf c}{\bu} \\
    \afi{\exst x^\sigma A(x)}{s,\br}{\bu} 
    	& \eqdef & \exists x \tin s \, \forall \bv \tin \bu \afi{A(x)}{\br}{\bv} 
\end{array}}
\]
\end{Definition}

Let us now show that this new presentation is indeed equivalent to the original presentation, in the sense that the formulas which are witnessable via the original presentation are precisely the formulas witnessable by the alternative presentation.

\begin{Definition} \label{up-down-constructions} Define four constructions to convert between elements of up and down types
\[
\begin{array}{ccc}
	\pup{(\cdot)}{A} \colon \typ^+_A \to \ty^+_A  
		& \hspace{10mm} & \pdown{(\cdot)}{A} \colon \ty^+_A \to \typ^+_A \\[2mm]
	\nup{(\cdot)}{A} \colon \typ^-_A \to \ty^-_A 
		& & \ndown{(\cdot)}{A} \colon \ty^-_A \to \typ^-_A
\end{array}
\]
as follows:
\[
{\arraycolsep=2pt
\def\arraystretch{1.4}
\begin{array}{rclcrcl}
    \pup{s}{(\st^\sigma(z))} & \eqdef & s 
    	& \hspace{15mm} & 
	\pup{(\br,\bt)}{(A \wedge B)} & \eqdef & \pup{\br}{A} , \pup{\bt}{B} \\
    \nup{\varepsilon}{(\st^\sigma(z))} & \eqdef & \varepsilon 
    	& &
	\nup{(\bu,\bv)}{(A \wedge B)} & \eqdef & \nup{\bu}{A} , \nup{\bv}{B} \\
    \pdown{s}{(\st^\sigma(z))} & \eqdef & s 
    	& &
	\pdown{(\br,\bt)}{(A \wedge B)} & \eqdef & \pdown{\br}{A} , \pdown{\bt}{B} \\
    \ndown{\varepsilon}{(\st^\sigma(z))} & \eqdef & \varepsilon
    	& &
	\ndown{(\bu,\bv)}{(A \wedge B)} & \eqdef & \ndown{\bu}{A} , \ndown{\bv}{B} \\[2mm]
    \pup{\br}{(\forall x^\sigma A)} & \eqdef & \pup{\br}{A}
    	& &
	\pup{\br}{(\exists x^\sigma A)} & \eqdef & \pup{\br}{A} \\
    \nup{\bu}{(\forall x^\sigma A)} & \eqdef & \nup{\bu}{A}
    	& &
	\nup{\bu}{(\exists x^\sigma A)} & \eqdef & \{ \nup{\bv}{A} : \bv \in \bu \} \\
    \pdown{\br}{(\forall x^\sigma A)} & \eqdef & \pdown{\br}{A}
    	& &
	\pdown{\br}{(\exists x^\sigma A)} & \eqdef & \pdown{\br}{A} \\
    \ndown{\bu}{(\forall x^\sigma A)} & \eqdef & \ndown{\bu}{A}
    	& &
	\ndown{\bu}{(\exists x^\sigma A)} & \eqdef & \{ \ndown{\bv}{A} : \bv \in \bu \} \\[2mm]
    \pup{\bdf}{(\allst x^\sigma A)} & \eqdef & \{ \lambda c^\sigma . \pup{(\bdf c)}{A} \}
    	& &
	\pup{(s,\br)}{(\exst x^\sigma A)} & \eqdef & s, \pup{\br}{A} \\
    \nup{(c, \bu)}{(\allst x^\sigma A)} & \eqdef & c, \nup{\bu}{A}
    	& &
	\nup{\bu}{(\exst x^\sigma A)} & \eqdef & \{ \nup{\bv}{A} : \bv \in \bu \} \\
    \pdown{\bdf}{(\allst x^\sigma A)} & \eqdef & \lambda c^\sigma . \pdown{(\bdf \apl{c})}{A}
    	& &
	\pdown{(s,\br)}{(\exst x^\sigma A)} & \eqdef & s, \pdown{\br}{A} \\
    \ndown{(c,\bu)}{(\allst x^\sigma A)} & \eqdef & c, \ndown{\bu}{A}
    	& &
	\ndown{\bu}{(\exst x^\sigma A)} & \eqdef & \{ \ndown{\bv}{A} : \bv \in \bu \} \\[2mm]
    \pup{(\bdf,\bdg)}{(A \to B)} & \eqdef &
    \multicolumn{5}{l}{
		\{ \lambda \br^{\ty^+_A} . \pup{(\bdf(\pdown{\br}{A}))}{B} \},
		\{ \lambda \br^{\ty^+_A} \lambda \bu^{\ty^-_B} . 
			\{ \nup{\bv}{A} \; : \; \bv \in \bdg \pdown{\br}{A} \ndown{\bu}{B} \} \} } \\
    \nup{(\br, \bu)}{(A \to B)} & \eqdef &
    \multicolumn{5}{l}{
    \pup{\br}{A}, \nup{\bu}{B}} \\
    \pdown{(\bdf,\bdg)}{(A \to B)} & \eqdef &
    \multicolumn{5}{l}{
		\lambda \br^{\typ^+_A} . \pdown{(\bdf \apl{\pup{\br}{A}})}{B},
		\lambda \br^{\typ^+_A} \lambda \bu^{\typ^-_B} . 
			\{ \ndown{\bv}{A} \; : \; \bv \in \bdg \apl{\pup{\br}{A}, \nup{\bu}{B}} \} } \\
    \ndown{(\br, \bu)}{(A \to B)} & \eqdef &
    \multicolumn{5}{l}{
    \pdown{\br}{A}, \ndown{\bu}{B} }
    
\end{array}}
\]
\end{Definition}

Let us illustrate this definition with our example $A \equiv \allst n^\N \exst m^\N ( n = m )$ with $\ty^+_A \equiv (\N \to \N^*)^*$. Given $t^{\ty^+_A} \equiv \{ \lambda n.\{ n \}, \lambda n.\{ n+1 \} \}$, we have that $\pdown{t}{A} \equiv \lambda n. \{ n, n+1 \}$. Intuitively, we can see here that $\pdown{(\cdot)}{A}$ works as a ``flattening" operation, inductively turning sets of functions (which produce sets) into a single function producing the union of the image sets.

\begin{Theorem} \label{main-thm} We have that
\begin{enumerate}
	\item $\hfi{A}{\br}{\nup{\bu}{A}} \vdash_{\HAomega} \afi{A}{\pdown{\br}{A}}{\bu}$,
	\item $\afi{A}{\br}{\ndown{\bu}{A}} \vdash_{\HAomega} \hfi{A}{\pup{\br}{A}}{\bu}$.
\end{enumerate}
\end{Theorem}
\Proof
We prove the two points by a simultaneous induction on the logical form of $A$. The only non-trivial cases are implication $A \to B$ and the universal quantifier $\allst x^\sigma A$: \\[2mm]
Implication 1.:
\[
\begin{array}{lcl}
\hfi{A \to B}{\bdf,\bdg}{\nup{(\br,\bu)}{(A \to B)}} 
	& \equiv & \hfi{A \to B}{\bdf,\bdg}{\pup{\br}{A},\nup{\bu}{B}} \\[2mm]
    	& \equiv & \forall \bv \tin \bdg\apl{\pup{\br}{A},\nup{\bu}{B}} \hfi{A}{\pup{\br}{A}}{\bv} 
			\to \hfi{B}{\bdf\apl{\pup{\br}{A}}}{\nup{\bu}{B}} \\[2mm]
    	& \IHarrow & 
		\forall \bv \tin \bdg\apl{\pup{\br}{A},\nup{\bu}{B}} \afi{A}{\br}{\ndown{\bv}{A}} 
			\to \afi{B}{\pdown{(\bdf\apl{\pup{\br}{A}})}{B}}{\bu} \\[2mm]
    	& \equiv & 
		\afi{A \to B}{\pdown{(\bdf,\bdg)}{(A \to B)}}{\br,\bu}
\end{array}
\]
Implication 2.: Let $\tilde{\bdf} \eqdef \{ \lambda \br^{\ty^+_A} . \pup{(\bdf(\pdown{\br}{A}))}{B} \}$ and $\tilde{\bdg} \eqdef \{ \lambda \br^{\ty^+_A} \lambda \bu^{\ty^-_B} .  \{ \nup{\bv}{A} \; : \; \bv \in \bdg \pdown{\br}{A} \ndown{\bu}{B}) \} \}$. Then
\[
\begin{array}{lcl}
\afi{A \to B}{\bdf,\bdg}{\ndown{(\br,\bu)}{(A \to B)}} 
	& \equiv & \afi{A \to B}{\bdf,\bdg}{\pdown{\br}{A},\ndown{\bu}{B}} \\[2mm]
    	& \equiv & \forall \bv \tin \bdg \pdown{\br}{A} \ndown{\bu}{B} \afi{A}{\pdown{\br}{A}}{\bv} 
			\to \afi{B}{\bdf \pdown{\br}{A}}{\ndown{\bu}{B}} \\[2mm]
    	& \stackrel{\small{\textup{IH}}}{\Rightarrow} & 
		\forall \bv \tin \bdg \pdown{\br}{A} \ndown{\bu}{B} \hfi{A}{\br}{\nup{\bv}{A}} 
			\to \hfi{B}{\pup{(\bdf \pdown{\br}{A})}{B}}{\bu} \\[2mm]
    	& \equiv & 
		\forall \bv \tin \tilde{\bdg} \apl{\br,\bu} \hfi{A}{\br}{\bv} 
			\to \hfi{B}{\tilde{\bdf} \apl{\br}}{\bu} \\[2mm]
    	& \equiv & 
		\hfi{A \to B}{\pup{(\bdf,\bdg)}{(A \to B)}}{\br,\bu}
\end{array}
\]
Universal (standard):
\[
\begin{array}{lcllcl}
	& & \hfi{ \allst x^\sigma A}{\bdf}{\nup{(c,\bu)}{(\allst x^\sigma A)}} & \hspace{20mm}
	& & \afi{\allst x^\sigma A}{\bdf}{\ndown{(c,\bu)}{(\allst x^\sigma A)}} \\[2mm]
	& \equiv & \hfi{\allst x^\sigma A}{\bdf}{c, \nup{\bu}{(A)}} &
	& \equiv & \afi{\allst x^\sigma A}{\bdf}{c, \ndown{\bu}{(A)}} \\[2mm]
    	& \equiv & \hfi{A[c/x]}{\bdf\apl{c}}{\nup{\bu}{A}} &
	    & \equiv & \afi{A[c/x]}{\bdf c}{\ndown{\bu}{A}} \\[2mm]
    	& \stackrel{\small{\textup{IH}}}{\Rightarrow} & 
		\afi{A[c/x]}{\pdown{(\bdf\apl{c})}{A}}{\bu} &
		& \stackrel{\small{\textup{IH}}}{\Rightarrow} & 
		\hfi{A[c/x]}{\pup{(\bdf c)}{A}}{\bu} \\[2mm]
    	& \equiv & 
		\afi{\allst x^\sigma A}{\pdown{\bdf}{(\allst x^\sigma A)}}{c,\bu} &
		& \equiv & 
		\hfi{\allst x^\sigma A}{\pup{\bdf}{(\allst x^\sigma A)}}{c,\bu}
\end{array}
\]
We leave the other simple cases to the reader.
\Done

\begin{Corollary} A principle is witnessable by the original Herbrand functional interpretation (Definition \ref{def-hfi}) iff it is witnessable by the alternative presentation of the Herbrand functional interpretation (Definition \ref{def-afi}). 
\end{Corollary}

\Proof Assume the original Herbrand functional interpretation of $A$ is witnessable, i.e.\ we have a term $\br^{\ty^+_A}$ such that $\forall \bu^{\ty^-_A} \hfi{A}{\br}{\bu}$. In particular we have $\forall \bu^{\typ^-_A} \hfi{A}{\br}{\nup{\bu}{A}}$ and, by Theorem \ref{main-thm} (1), $\forall \bu^{\typ^-_A} \afi{A}{\pdown{\br}{A}}{\bu}$. So that $\pdown{\br}{A}$ can be seen to be a witness for the alternative Herbrand functional interpretation of $A$. The converse is proved similarly using Theorem \ref{main-thm} (2).
\Done


Since in $\afi{A}{\br}{\bu}$ the witnesses $\br$ need not be of finite set type, the monotonicity property as formulated in Lemma~\ref{monotonicity} does not even make sense for $\afi{A}{\br}{\bu}$. We conclude by showing that a monotonicity property similar to (the stronger) Lemma~\ref{monotonicity2} also holds for the alternative presentation of the Herbrand functional interpretation.

\begin{Definition} A finite type $\sigma$ is called a $*$-type if it is either of the form $\rho^*$, for some finite type $\rho$, or of the form $\rho \to \tau$ with $\tau$ being a $*$-type. Define a partial order $x \preceq_\sigma x'$ on $*$-types inductively as follows:
\[
{\arraycolsep=2pt
\def\arraystretch{1.4}
\begin{array}{rcl}
s \preceq_{\sigma^*} s' & \eqdef & s \subseteq_{\sigma^*} s' \\
f \preceq_{\sigma \to \rho} f' & \eqdef & \forall x^{\sigma} \left( fx \preceq_\rho f'x \right) \\
\end{array}}
\]
If $\bt,\bt'$ are tuples of terms of types $\mathbf{\sigma}$, then we write
\[
\bt \preceq_{\mathbf{\sigma}} \bt' \, \eqdef \, t_1 \preceq_{\sigma_1} t'_1 \wedge \ldots \wedge t_n \preceq_{\sigma_n} t'_n.
\]
\end{Definition}

Intuitively, $t \preceq_\sigma t'$ is the usual subset relation when $t$ is a finite set, but it is defined inductively and pointwise when $t$ is a functional. 

\begin{Lemma} When $\mathbf{\sigma}$ is a tuple of positive down type $\mathbf{\sigma} \equiv \typ^+_A$, we will write $\br \preceq_A \br'$ instead of $\br \preceq_{\typ^+_A} \br'$. Noting that $\typ^+_A$ is a sequence of $*$-types, we have:
\[
{\arraycolsep=2pt
\def\arraystretch{1.5}
\begin{array}{rcl}
    s \preceq_{\st^\sigma(z)} s' 
    	& \equiv & s \subseteq_{\sigma^*} s' \\
    \br, \bt \preceq_{A \wedge B} \br', \bt' 
    	& \equiv & (\br \preceq_A \br') \wedge (\bt \preceq_B \bt') \\
    \bdf, \bdg \preceq_{A \to B} \bdf', \bdg' 
    	& \equiv & \forall \br (\bdf \br \preceq_B \bdf' \br ) \wedge
			\forall \br, \bu (\bdg \br \bu \subseteq_{(\typ^-_A)^*} \bdg' \br \bu)  \\
    \br \preceq_{\forall x^\sigma A} \br' 
    	& \equiv & \br \preceq_{A} \br' \\
    \br \preceq_{\exists x^\sigma A} \br' 
    	& \equiv & \br \preceq_{A} \br' \\
    \bdf \preceq_{\allst x^\sigma A} \bdf' 
    	& \equiv & \forall c^\sigma (\bdf c \preceq_{A} \bdf' c) \\
    s, \br \preceq_{\exst x^\sigma A} s', \br' 
    	& \equiv & (s \subseteq_{\sigma^*} s') \wedge (\br \preceq_{A} \br')
\end{array}}
\]
\end{Lemma}

The alternative presentation of the Herbrand functional interpretation satisfies the following monotonicity property:

\begin{Lemma}[Monotonicity property] \label{mon-strong} 
$\afi{A}{\br}{\bu} \wedge \br \preceq_A \br' \vdash_{\HAomega} \afi{A}{\br'}{\bu}$.
\end{Lemma}
\Proof
By induction on $A$. Here we prove only the case of implication. Assume that $\afi{A \to B}{\bdf,\bdg}{\bu,\br}$ and $\bdf,\bdg \preceq_{A \to B} \bdf',\bdg'$. We have $\bdf\br \preceq_B \bdf'\br$ and $\bdg\br\bu \subseteq_{(\typ^-_A)^*} \bdg'\br\bu$ and hence
\[
\afi{A \to B}{\bdf,\bdg}{\bu,\br}
\,\equiv\, \forall \bv \tin \bdg \br \bu \afi{A}{\br}{\bv} \to \afi{B}{\bdf \br}{\bu}
\,\IHarrow\, \forall \bv \tin \bdg' \br \bu \afi{A}{\br}{\bv} \to \afi{B}{\bdf' \br}{\bu}\,\equiv\, \afi{A \to B}{\bdf',\bdg'}{\bu,\br}. 
\]
\Done

\bibliographystyle{plain} 

\bibliography{references}

\end{document}